\documentclass{aa}  

\usepackage{epsf}
\usepackage{txfonts}
\usepackage{color}

\usepackage{natbib}
\bibpunct{(}{)}{;}{a}{}{,} 

\usepackage{graphicx}
\usepackage{hyperref}

\def\DM{\ensuremath{D\hspace{-0.1ex}M}}

\def\WNEs{\mbox{WNE-s}}
\def\WNEw{\mbox{WNE-w}}
\def\*{$^*$}

\begin{document}

\title{The Galactic WN stars revisited}

\subtitle{Impact of Gaia distances on fundamental stellar parameters}

\author{W.-R. Hamann\inst{1}
  \and
G. Gr\"afener\inst{2}
  \and
A. Liermann\inst{\ref{inst1},\ref{inst3}}
  \and
R. Hainich\inst{\ref{inst1}}
  \and
A.A.C. Sander\inst{\ref{inst1},\ref{inst4}}
  \and
T. Shenar\inst{\ref{inst1},\ref{inst5}}
  \and
V. Ramachandran\inst{\ref{inst1}}
  \and
H.~Todt\inst{\ref{inst1}}
  \and
L.M. Oskinova\inst{\ref{inst1},\ref{inst6}}
}

\institute{Institut f\"ur Physik und Astronomie, Universit\"at
Potsdam, Karl-Liebknecht-Str. 24, 14476 Potsdam, Germany \label{inst1}\\ 
\email{wrh@astro.physik.uni-potsdam.de} 
   \and
Argelander-Institut f\"ur Astronomie der Universit\"at Bonn, Auf dem
H\"ugel 71, 53121, Bonn, Germany \label{inst2}
   \and
Leibniz-Institut f\"ur Astrophysik (AIP), An der Sternwarte 16,
14482 Potsdam, Germany \label{inst3}
   \and
Armagh Observatory and Planetarium, College Hill, Armagh, 
BT61 9DG, Northern Ireland\label{inst4}
   \and
Institute of Astrophysics, KU Leuven, Celestijnenlaan 200 D, 3001
Leuven, Belgium \label{inst5}
   \and
Kazan Federal University, Kremlevskaya Ul. 18, Kazan, Russia \label{inst6}
}

\date{Received ...; accepted ...}

 
\abstract
{Comprehensive spectral analyses of the Galactic Wolf-Rayet stars of
the nitrogen sequence (i.e.\ the WN subclass) have been performed in a
previous paper. However, the distances of these objects were poorly
known. Distances have a direct impact on the ``absolute'' parameters,
such as luminosities and mass-loss rates. The recent Gaia Data Release 
(DR2) of trigonometric parallaxes includes
nearly all WN stars of our Galactic sample. In the present paper,  we
apply the new distances to the previously analyzed Galactic WN stars
and rescale the results accordingly. On this basis, we present a
revised catalog of 55 Galactic WN stars with their stellar and
wind parameters.  The correlations between mass-loss rate and
luminosity show a large scatter, for the hydrogen-free WN stars as well
as for those with detectable hydrogen. The slopes of the $\log L - \log
\dot{M}$ correlations are shallower than found previously.  The
empirical Hertzsprung-Russell diagram (HRD) still shows the previously
established dichotomy between the hydrogen-free early WN subtypes that 
are located on the hot side of the zero-age main sequence (ZAMS), and the 
late WN subtypes, which show hydrogen and reside mostly at cooler
temperatures than the ZAMS (with few exceptions).  However, with the
new distances, the distribution  of stellar luminosities became  more
continuous than obtained previously. The hydrogen-showing stars
of late WN subtype are still found to be typically more luminous than 
the hydrogen-free early subtypes, 
but there is a range of luminosities where both subclasses
overlap. The empirical HRD of the Galactic single WN stars is compared
with recent evolutionary tracks. Neither these single-star
evolutionary models nor binary scenarios can provide a fully
satisfactory explanation for the parameters of
these objects and their location in the HRD. 
}

\keywords{stars: mass-loss
  -- stars: winds, outflows -- stars: Wolf-Rayet -- stars: atmospheres
  -- stars: evolution -- stars: distances }

\maketitle


\section{Introduction}

Massive stars play an essential role in the Universe.
With their feedback of ionizing photons and their stellar mass loss, these stars
govern the ecology of their extended neighborhood. Their final
gravitational collapse is probably in most cases accompanied by a
supernova explosion or even  gamma-ray burst. In case a close pair of
compact objects (neutron stars or black holes) is left over, these
compact remnants will spiral in over gigayears until they finally merge 
with a spectacular
gravitational wave event as those detected in the recent years.  
 
Many massive stars will pass through Wolf-Rayet (WR) stages in advanced
phases of their evolution. However, these last stages of massive-star
evolution are yet poorly understood.  Major uncertainties are induced
by the lack of knowledge in various aspects:  the stellar mass loss in
the O-star phase and in the WR phase(s)
\citep[e.g.,][]{mokiem2007,hainich+2015};  eruptive mass-loss
\citep[e.g., luminous blue variables;][]{smith+owocki2006}; 
mass loss in the red supergiant stage \citep{Vanbeveren+2007};  angular
momentum loss; internal mixing processes, including rotationally
induced mixing \citep[e.g.,][]{maeder2003}; binary fraction
\citep{sana2017};  close-binary evolution with mass exchange
\citep[e.g.,][]{marchant+2016}; common-envelope evolution
\citep{paczynski1976}; and stellar mergers.  Complicating things even
more, all these issues depend on the metallicity  of the respective
stars.  
 
In this situation, it is crucial to establish empirical constraints. 
We previously analyzed a comprehensive sample of Galactic WN stars  
\citep[][hereafter Paper\,I]{Hamann+2006}. This homogeneous set of
spectral  analyses was based on the Potsdam Wolf-Rayet (PoWR)  model
atmospheres\footnote{\url{http://www.astro.physik.uni-potsdam.de/PoWR/}}. 
This state-of-the-art non-LTE code solves the radiative transfer in a
spherically expanding atmosphere and accounts not only for complex
model atoms, but also for iron line blanketing and for wind clumping
\citep{graefener2002, hamann2004}. 

A basic disadvantage of the Galactic WN sample was the large uncertainty
of the stellar distances. While the basic spectroscopic parameters
(especially the stellar temperature $T_\ast$ and the 
``transformed radius'' $R_\mathrm{t}$ ; see
Sect.\,\ref{sect:parameters}) do not depend on the adopted distance, the
``absolute'' parameters (luminosity $L$, mass-loss rate $\dot{M}$) do. 

This situation fundamentally improved with the recent Gaia Data
Release\,2 (DR2). Fortunately, WR spectra closely follow a
scaling invariance, which makes it possible to rescale the results from
Paper\,I to the new Gaia distances without redoing the whole spectral
analyses. This is the core of the present paper. The same task has already been 
carried out for the Galactic WC and WO stars \citep{Sander+2019}. 
Naturally, our analyses of WR stars in the Magellanic
Clouds \citep{Hainich+2014,hainich+2015, Shenar+2016, Shenar+2018} or
in M31 \citep{Sander+2014} do not suffer from distance uncertainties. 

The rest of the paper is structured as follows. In Sect.\,2 we introduce
our sample and review the binary status. In Section\,3 we adapt the Gaia DR2
parallaxes and distances for our sample, and investigate the impact of the
revised distances on the WN star positions in the Galaxy and on their
absolute visual magnitudes. Section\,4 describes the rescaling of the
stellar parameters from Paper\,I, while the results of this procedure
are visualized in Sect.\,5; especially, the updated Hertzsprung-Russell 
diagram (HRD) is discussed with regard to the evolutionary connections. A
summary is given in the last section (Sect.\,\ref{sect:summary}).


\section{The sample of WN stars}
\label{sect:binarity}

The present study relies on the same stars as Paper\,I (cf.\ 
Table\,\ref{table:parameters}). 
The list of objects analyzed in Paper\,I comprises the vast majority of 
Galactic WN stars that can be observed in visual light since they are
not obscured too much by interstellar absorption. We left out those WN
stars that show composite spectra, typically as WR+OB binaries. The
concern at this point is that composite spectra need to be analyzed as
such; neglecting their composite nature would compromise the results.
Analyzing composite spectra is of course also manageable
\citep[see, e.g.,][]{Shenar+2018, shenar+2017, Shenar+2016}, but requires
more and better observational data and has been beyond the scope of Paper\,I.  

In order to make sure that the sample does not contain composite spectra
with a significant contribution from the non-WR component, we 
checked the literature that has appeared since publication of Paper\,I.
Below, we comment on the binary status of the individual stars in cases in which 
new evidence has been published. 

\medskip\noindent\textbf{WR1} is still considered to be single; the
observed polarization is attributed to corotating interaction regions
(CIRs) rather than to wind asymmetry 
\citep{St-Louis2013}. 

{
\noindent\textbf{WR2} has been extensively observed by
\cite{Chene+2019}, who did not find any evidence for binarity. These authors
identified only a small contribution (5\%) to the optical spectrum by a
B-type star in the background with a projected distance of 0\farcs25.  
}

\noindent\textbf{WR6} is still considered to be single; the polarimetric and
photometric variability was rediscussed by \cite{St-Louis+2018} 
and attributed to CIRs. 

\noindent\textbf{WR12} is clearly a binary with colliding winds;
however, no traces of the putative OB-type companion could be identified in the 
spectrum \citep{fahed+moffat2012}, indicating that its contribution is
minor.

\noindent\textbf{WR22} is clearly a WR+O binary; \cite{Rauw+1996}
estimated that the O-type companion is about eight times fainter than the WR
primary (at 5500\,\AA). Thus, neglecting the contribution of the O star should 
introduce only a limited bias on the spectral analysis. 
 
\noindent\textbf{WR25} shows an X-ray light curve with a period of 208\,d, 
confirming a WR+O colliding wind system \citep{pandey+2014}. Its  
eccentric orbit has been established
first by \cite{Gamen+2008}, who claimed minimum masses of 75 +
27\,$M_\odot$, but did not report a brightness ratio.
The distance to WR\,25 has been revised significantly by Gaia, from
\DM = 12.55\,mag (Paper\,I) to 11.5\,mag
{ corresponding to 2.0\,kpc, which is nearer than the bulk of
the Carina nebula}.

\noindent\textbf{WR40} is unusually faint in X-rays, 
{ 
in contrast to the expectation that it could be
}a colliding-wind binary \citep{Gosset+2005}.
  
\noindent\textbf{WR44} shows variability that is typical for CIRs rather
than indicating binarity \citep{Chene+2011}.


\noindent\textbf{WR47} was excluded from the sample of Paper\,I because
of suspected composite spectrum. Meanwhile, it has been studied by 
\cite{fahed+moffat2012} as a colliding-wind system. O-star features
could not be identified in the spectrum, implying that the
contribution of this component to the total light is small. Hence, we
actually could have kept the spectrum in our single-star analysis. 
WR\,47 is a runaway  \citep{tetzlaff+2011}.

\noindent\textbf{WR66} 
{ 
has 
}
a negative parallax measurement in Gaia DR2,
probably because the visual companion which is separated by 0.4\arcsec
and 1.05\,mag fainter was 
{ interfering with the astrometry. 
}


\noindent\textbf{WR78} shows no evidence of binarity, although
\cite{skinner+2012} note that its X-rays include a hot plasma 
component ``at the high end of the range for WN stars''.

\noindent\textbf{WR89} is a 
thermal radio source \citep{montes+2009}, supporting the assumption 
that it is single. 

\noindent\textbf{WR105} shows 
non-thermal radio emission \citep{montes+2009}, which could be indicative
of colliding winds, but there are no other signs of binarity. 

\noindent\textbf{WR107} got a negative parallax measurement in
Gaia DR2 for unknown reasons. 


\noindent\textbf{WR123} has been extensively monitored with MOST
\citep{Lefevre+2005} and showed  no stable periodic signals that could
be attributed to orbital modulation (periods above one day). A
persistent signal with  about 9.8\,h period is likely related to 
pulsational instabilities as claimed by \cite{Dorfi+2006}, although
their stability analysis is based on the wrong assumption of a
hydrogen-rich envelope.  

\noindent\textbf{WR124} has been reported by \cite{Moffat+1982} to show
radial velocity variations. These authors suggested that the star might be a
binary hosting a compact object. However, \cite{Marchenko+1998-wn8}
could not confirm these radial velocity variations. A period search in
the photometric data  from {\sc hipparcos} also remained without
significant detection  \citep{Marchenko+1998-hipparcos}. The star is 
very faint in X-rays ($L_{\rm X}\sim 10^{31}$\,erg\,s$^{-1}$).
Nevertheless, Toal\'a et al.\ (2018, in press) showed that a hypothetical compact
companion might be hidden deep in the wind. So far, WR\,124 must be
considered as being single.

\noindent\textbf{WR130} got a negative parallax measurement in
Gaia DR2 for unknown reasons. 

\noindent\textbf{WR134} shows spectral variability with a period of 
2.25\,d, which has been revisited in a long observational campaign 
by \cite{Aldoretta+2016} and attributed to CIRs rather than to binarity. 

\noindent\textbf{WR136} is a runaway star \citep{tetzlaff+2011}.


\noindent\textbf{WR147}  is obviously a binary system with colliding
winds. {\em Chandra} resolved a pair of  X-ray sources
\citep{zhekov+park2010}. High-resolution radio observations also
resolved two components, the southern thermal source WR147S (the WN8
star), and a northern non-thermal source WR 147N with a mutual
separation of 0\farcs57 \citep{Abbott+1986, Moran+1989,
Churchwell+1992, Contreras+1996, Williams+1997, Skinner+1999}. The
binary system was also spatially resolved in the infrared and optical
range with a  separation of 0\farcs64 \citep{Williams+1997,
Niemela+1998}. According to the latter work, the companion is by a
factor 7.3 fainter in the visual than the WN8 primary. Hence, our
single-star spectral analysis might be slightly biased by contributions
from the secondary and the colliding-wind zone. 
Gaia DR2 obtained a negative parallax for this object, presumably 
because the multiplicity of the light source has 
irritated the measurements.  

\noindent\textbf{WR148} had been suspected to host a compact
companion \citep{Marchenko+1996}. However, \cite{munoz+2017} identified
O-star features in the composite spectrum, and thus established a binary
system with components classified as WN7h and O4-6 in a 4.3\,d orbit.  
While the brightness ratio was not determined, we must be aware that
our single-star analysis might be somewhat biased by the contribution
of the companion. The Gaia DR2 catalog reports a negative parallax, but it
seems unlikely that is was the close-binary nature of the object that 
has irritated the measurements. With a height of about 800\,pc above the
Galactic plane, WR\,148 is an extreme runaway object. 


\noindent\textbf{WR156} shows radio emission with a thermal and a
non-thermal contribution \citep{montes+2009}, but otherwise there are no
indications of binarity. 


In those figures that indicate results of our spectral analysis, we
mark the confirmed WR+OB binaries by encircling their respective symbol. 
Runaway stars are marked in the same way since their fast motion is
most likely the result of former binarity, after the companion exploded
and became gravitationally unbound.  

We emphasize that all distance-independent results from the spectral
analyses in Paper\,I are retained for the present study -- notably the 
stellar temperatures,
``transformed radii'' (cf.\ Sect.\,\ref{sect:parameters}), and the
detectability  of atmospheric hydrogen. Paper\,I provides 
a discussion of the error margins. The spectral types
given in Column (2) of  Table\,\ref{table:parameters} are copied from
Paper\,I as well.  We note that some of the stars with ``early'' subtype
numbers are designated as ``(WNL)'' in parentheses, while some of the
stars with subtype numbers as high as 7 or 8 are still classified in
parantheses  as ``(WNE-w)'', i.e., as early-type WN with weak lines. In
Paper\,I (end of section\,3) we argued that such assignment corresponds
better to the spectral appearance. Since the current paper only
updates the distance-dependent quatities, we keep these spectral type
assignments throughout the present paper, including the symbol coding
in the diagrams.   

\section{Gaia distances for the Galactic WN stars}
\label{sect:distances}

One of 
the main objectives of the Gaia space craft is to measure stellar
parallaxes all over our Galaxy \citep{prusti+2016}. The original Gaia
DR2
catalog\footnote{http://vizier.u-strasbg.fr/viz-bin/VizieR?-source=I/345} 
provides parallaxes for 1.8\,billion stars, including all our targets
except of WR\,2 and WR\,63.

Ideally, the distance $d$ to a particular object is just the reciprocal 
of its parallax $d = \varpi^{-1}$. However, the measured parallax has 
a statistical error. As consequence, the most likely value for the distance
is not exactly the reciprocal parallax.  \citet{Bailer-Jones+2018}
provided a version of the Gaia DR2 catalog in which the distances are
derived from the parallaxes with the help of a Bayesian
approach. This catalog also provides error margins for the distance
that correspond to a confidence level that is equivalent to $1\sigma$ in a
Gaussian distribution. 

We decided to retrieve the distances of our targets from the Bailer-Jones
et al.\ catalog, although we are aware that this approach is based on a
general model of the stellar density in our Galaxy. 
\citet{Bailer-Jones+2018} gives positive distances even for those   
five stars of our sample (WR\,66, 107, 130, 147, and 148)
for which the original Gaia measurements resulted in 
negative parallaxes. In these cases the Bailer-Jones
et al.\  distances mainly reflect the adopted Galactic model, and we
disregard these values. We checked that they would lead to spurious
results for the stellar parameters. Moreover, there is one star in the
sample, WR\,115, for which Gaia gives an amazingly small distance of
570\,pc, but with a huge $1\sigma$ error from 335 to 2006\,pc; this is by
far the largest error margin within the sample. This measurement leads to
implausible stellar parameters, and is also disregarded in the rest of
this paper.   

\begin{figure}[tb]
\centering
\epsfxsize=\columnwidth
\mbox{\epsffile{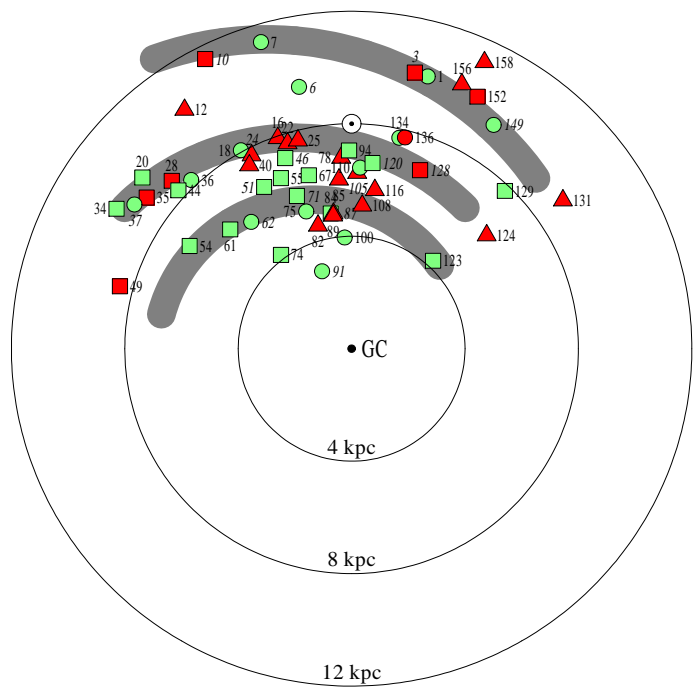}}
\caption{Galactic position of the WN stars based on Gaia distances. 
The meaning of the different symbols is explained in the caption of 
Fig.\,\ref{fig:mvcalib}. The labels refer to the WR catalog numbers. 
The Sun ($\odot$) and the Galactic Center (GC) are indicated. The
WN stars trace the Galactic spiral arm structure as indicated
schematically \citep[after][]{vallee2016} by the shaded arcs for the 
Perseus (outmost), Carina-Sagittarius (middle), and Crux-Centaurus
(innermost) arm, respectively. 
}
\label{fig:gal-KOS}
\end{figure}

As mentioned above, the distances from the \cite{Bailer-Jones+2018}
version of the Gaia catalog are obtained with a Bayesian appoach, for
which a specific model for the stellar density in our Galaxy had been
adopted. This {\em prior} might not necessarily apply to our targets;
for example, WR stars might be concentrated in the spiral arms. However,
the construction of a special prior for WR stars bears the danger that
the results would be biased toward the expectations. Therefore we
refrain from such attempts. 

The Bayesian approach leads systematically to lower distances than the 
reciprocal parallax. Fortunately, this effect only becomes noticeable at
largest distances, and thus affects only a few of our targets. For a
Bayesian distance modulus of 14\,mag the difference is about 0.5\,mag and thus
comparable to the typical 1$\sigma$ uncertainty; for the farthest
star in our sample (WR\,49), the inverse parallax yields a distance
modulus of 16.1\,mag instead of 15.0\,mag from the Bayesian approach.     
Our conclusions do not depend critically on these uncertainties, as we 
have checked. 

For a couple of our targets, the Gaia\,DR2 catalog gives a
non-zero {\tt astrometric\_excess\_noise} which indicates a poor fit to
the astrometric measurements \citep{Arenou+2018,Lindegren+2018}.
Such poor astrometric solution might compromise the parallax
measurement, although this is not always reflected by especially large
error margins. We consider a non-zero {\tt astrometric\_excess\_noise}
as a warning that the trigonometric distance might be less reliable
and indicate the corresponding stars in the figures by printing their names 
in slanted font. In Table\,\ref{table:parameters} the letter between
Columns (9) and (10) coding for the availability of a Gaia distance is a
lower case ``g'' in these cases. However, the results shown in the rest
of the paper do not
give the impression that the flagged stars are outliers regarding their
distance-dependent parameters, and thus do not 
substantiate doubts on their parallaxes.

The Galactic positions of our sample stars with Gaia distances, neglecting
their height over the Galactic plane, are plotted in Fig.\,\ref{fig:gal-KOS}.
As to be expected, they closely indicate the nearby spiral arms.

Essential for the present paper are the new Gaia distances, which
are given in Table\,\ref{table:parameters} 
Column (9) in form of the distance modulus, 
\begin{equation}
\DM = 5 \log~ (d/\mathrm{10pc})
,\end{equation}
flagged with the subsequent letter ``G'' when available, together with
the error margins as described above.  For the few stars without Gaia
distance, Table\,\ref{table:parameters} repeats the \DM\ values from
Paper\,I, and the subsequent arrow indicates whether this distance was
derived from some cluster or association membership
($\tiny\rightarrow$) or from the adopted subtype calibration of the
absolute visual magnitude ($\tiny\leftarrow$); see Paper\,I for
details. Anyhow, all stars  without Gaia distances are disregarded
in the rest of this paper. 


\def\WNEs{\mbox{WNE-s}}
\def\WNEw{\mbox{WNE-w}}
\def\*{$^*$}

\def\R{\,\raisebox{0.4ex}{\tiny$\rightarrow$}\,}
\def\l{\,\raisebox{0.4ex}{\tiny$\leftarrow$}\,}
\def\noobs{\multicolumn{3}{c}{-- no observation --}}
\def\comp{\multicolumn{3}{c}{-- composite spectrum --}}
\def\s{\rule[0mm]{0mm}{4.5mm}}
\def\a{$^a$}
\def\b{$^b$}
\def\c{$^c$}
\tabcolsep 4.3pt
\begin{table*}[!hbtp]
\caption[]{
Parameters of the Galactic single WN stars
\label{table:parameters}
}
\small
\begin{flushleft}
\begin{tabular}{ l l r r r r r l r c r l r r }
\hline\hline 
\s WR &
Spectral subtype &
$T_*$ &
$\log R_{\rm t}$ &
\multicolumn{1}{c}{$\varv_\infty$} &
$X_{\rm H}$ &
$E_{b-\varv}$ &
Law\a  &
\DM\c \hspace{5mm} $M_\varv$ &
$R_*$& log\,$\dot{M}^d$ &
log\,$L$ &
$\frac{{\dot M} \varv_{\infty}}{L/c}$ &
$M^e$ \\

&
&
[kK] &
$[R_{\odot}]$ &
\multicolumn{1}{c}{[km/s]} &
[\%] &
[mag]&
$R_V$  &
[mag]\ \ \ [mag] &
$[R_{\odot}]$ &
[$M_{\odot}$/yr] &
[$L_{\odot}$] &
&
[$M_{\odot}$] \\

   (1) & (2)           &(3)      & (4)   & \multicolumn{1}{c}{(5)}
                                             &  (6) & (7)&  (8) & (9) \hspace{0.6cm} (10)
                                                                                   & (11) & (12) & (13)& (14) &(15) \\
\hline 
\s 1    & WN4-s         & 112.2 & 0.3 & 1900 &  0 & 0.67 & S     & 12.5$^{+0.2}_{-0.2}$ G  -4.74 & 2.26 & -4.3 & 5.88 &  5.6 &  27\\[0.6mm]
   2    & WN2-w         & 141.3 & 0.5 & 1800 &  0 & 0.44 & C 3.0 & 12.0                 \R -2.43 & 0.89 & -5.3 & 5.45 &  1.7 &  16 \\[0.6mm]
   3    & WN3h-w        &  89.1 & 1.2 & 2700 & 20 & 0.35 & C 3.4 & 12.3$^{+0.3}_{-0.2}$ g  -3.13 & 2.48 & -5.4 & 5.56 &  1.4 &  17 / 15\\[0.6mm]
   6    & WN4-s         &  89.1 & 0.3 & 1700 &  0 & 0.12 & S     & 11.8$^{+0.3}_{-0.2}$ g  -5.33 & 3.25 & -4.2 & 5.79 &  9.4 &  23\\[0.6mm]
   7    & WN4-s         & 112.2 & 0.3 & 1600 &  0 & 0.53 & S     & 13.2$^{+0.4}_{-0.4}$ G  -3.62 & 1.26 & -4.8 & 5.36 &  5.9 &  13\\[0.6mm]
   10   & WN5ha-w       &  63.1 & 1.2 & 1100 & 25 & 0.58 & C 3.1 & 13.8$^{+0.4}_{-0.4}$ g  -5.08 & 6.93 & -5.2 & 5.83 &  0.5 &  25 / 23\\[0.6mm]
   12\b & WN8h + OB     &  44.7 & 1.0 & 1200 & 27 & 0.80 & C 3.7 & 13.9$^{+0.4}_{-0.3}$ G  -6.68 &16.38 & -4.3 & 5.98 &  3.4 &  31 / 30\\[0.6mm]
   16   & WN8h          &  44.7 & 0.9 &  650 & 25 & 0.55 & C 3.4&  12.1$^{+0.2}_{-0.2}$ G  -6.14 &11.56 & -4.6 & 5.72 &  1.5 &  21 / 19\\[0.6mm]
   18   & WN4-s         & 112.2 & 0.3 & 1800 &  0 & 0.75 & C 3.6 & 13.0$^{+0.4}_{-0.3}$ G  -5.36 & 2.82 & -4.1 & 6.11 &  5.0 &  38\\[0.6mm]
   20   & WN5-w         &  63.1 & 0.9 & 1200 &  0 & 1.28 & S     & 14.4$^{+0.3}_{-0.3}$ G  -5.06 & 6.89 & -4.5 & 5.84 &  2.6 &  25\\[0.6mm]

\s 21   & WN5 + O4-6    &  \comp \\[0.6mm]
   22\b & WN7h + O9III-V&  44.7 & 1.3 & 1785 & 44 & 0.35 & C 3.8 & 11.9$^{+0.2}_{-0.2}$ G  -7.17 &22.65 & -4.4 & 6.28 &  1.8 &  49 / 75\\[0.6mm]
   24   & WN6ha-w (WNL) &  50.1 & 1.35& 2160 & 44 & 0.24 & C 3.1 & 12.8$^{+0.3}_{-0.3}$ g  -7.34 &21.73 & -4.3 & 6.47 &  1.8 &  68 /114\\[0.6mm]
   25\b & WN6h-w+O (WNL)&  50.1 & 1.5 & 2480 & 53 & 0.63 & C 4.5 & 11.5$^{+0.1}_{-0.1}$ G  -6.98 &20.24 & -4.6 & 6.38 &  1.2 &  58 / 98\\[0.6mm]
   28   & WN6(h)-w      &  50.1 & 1.2 & 1200 & 20 & 1.20 & S     & 14.1$^{+0.4}_{-0.4}$ G  -6.08 &14.06 & -4.7 & 6.06 &  1.0 &  35 / 35\\[0.6mm]
   31   & WN4 + O8V     &  \comp \\[0.6mm]
   34   & WN5-w         &  63.1 & 0.8 & 1400 &  0 & 1.18 & S     & 14.7$^{+0.4}_{-0.3}$ G  -5.08 & 6.28 & -4.5 & 5.75 &  3.8 &  22\\[0.6mm]
   35   & WN6h-w        &  56.2 & 0.9 & 1100 & 22 & 1.15 & S     & 14.4$^{+0.3}_{-0.3}$ G  -5.29 & 7.34 & -4.7 & 5.69 &  2.0 &  20 / 18\\[0.6mm]
   36   & WN5-s         &  89.1 & 0.2 & 1900 &  0 & 1.00 & S     & 13.9$^{+0.4}_{-0.3}$ G  -4.44 & 1.79 & -4.3 & 5.30 & 23.6 &  12\\[0.6mm]
   37   & WN4-s         & 100.0 & 0.4 & 2150 &  0 & 1.63 & S     & 14.6$^{+0.4}_{-0.4}$ g  -5.20 & 3.37 & -4.2 & 6.05 &  6.1 &  34\\[0.6mm]

\s 40   & WN8h          &  44.7 & 0.7 & 650  & 23 & 0.40 & C 3.4 & 13.0$^{+0.3}_{-0.3}$ G  -6.88 &14.51 & -4.2 & 5.91 &  2.5 &  28 / 26\\[0.6mm]
   44   & WN4-w         &  79.4 & 0.8 & 1400 &  0 & 0.62 & C 3.6 & 14.1$^{+0.4}_{-0.3}$ G  -4.02 & 3.37 & -4.9 & 5.62 &  1.9 &  18\\[0.6mm]
   46   & WN3p-w        & 112.2 & 0.8 & 2300 &  0 & 0.30 & F 3.6 & 12.1$^{+0.2}_{-0.2}$ g  -2.56 & 1.36 & -5.4 & 5.42 &  1.8 &  14\\[0.6mm]
   47   & WN6 + O5V     &  \comp \\[0.6mm]
   49   & WN5(h)-w      &  56.2 & 1.0 & 1450 & 25 & 0.80 & S     & 15.0$^{+0.4}_{-0.3}$ G  -4.45 & 5.20 & -5.0 & 5.40 &  2.8 &  14\\[0.6mm]
   51   & WN4-w         &  70.8 & 0.9 & 1500 &  0 & 1.40 & S     & 12.9$^{+0.2}_{-0.2}$ g  -3.85 & 3.72 & -5.0 & 5.50 &  2.3 &  16\\[0.6mm]
   54   & WN5-w         &  63.1 & 0.9 & 1500 &  0 & 0.82 & S     & 14.3$^{+0.4}_{-0.4}$ G  -4.63 & 5.65 & -4.7 & 5.67 &  2.8 &  20\\[0.6mm]
   55   & WN7 (WNE-w)   &  56.2 & 0.8 & 1200 &  0 & 0.65 & C 3.6 & 12.5$^{+0.3}_{-0.3}$ G  -4.67 & 5.23 & -4.7 & 5.40 &  4.7 &  14\\[0.6mm]
   61   & WN5-w         &  63.1 & 0.7 & 1400 &  0 & 0.55 & C 2.9 & 13.8$^{+0.4}_{-0.4}$ G  -3.53 & 2.75 & -5.0 & 5.03 &  6.8 &   9\\[0.6mm]
   62   & WN6-s         &  70.8 & 0.4 & 1800 &  0 & 1.73 & S     & 13.5$^{+0.4}_{-0.4}$ g  -6.34 & 6.32 & -3.8 & 5.96 & 14.8 &  30\\[0.6mm]

\s 63   & WN7 (WNE-w)   &  44.7 & 1.1 & 1700 &  0 & 1.54 & C 3.1 & 12.2                 \l -5.67 & 11.2 & -4.6 & 5.65 &  5.3 &  20 \\[0.6mm]
   66   & WN8(h)        &  44.7 & 0.9 & 1500 &  5 & 1.00 & S     & 14.8                 \l -7.22 & 19.9 & -3.9 & 6.15 &  6.2 &  41 \\[0.6mm]
   67   & WN6-w         &  56.2 & 0.8 & 1500 &  0 & 1.05 & S     & 11.9$^{+0.4}_{-0.4}$ G  -4.03 & 3.73 & -4.8 & 5.11 &  8.7 &   9\\[0.6mm]
   71   & WN6-w         &  56.2 & 0.9 & 1200 &  - & 0.38 & F 2.5 & 12.5$^{+0.3}_{-0.3}$ g  -3.59 & 3.56 & -5.1 & 5.06 &  3.7 &   9\\[0.6mm]
   74   & WN7 (WNE-w)   &  56.2 & 0.7 & 1300 &  0 & 1.50 & S     & 13.6$^{+0.6}_{-0.5}$ G  -5.79 & 6.91 & -4.4 & 5.65 &  5.6 &  19\\[0.6mm]
   75   & WN6-s         &  63.1 & 0.6 & 2300 &  0 & 0.93 & S     & 12.7$^{+0.4}_{-0.3}$ G  -5.30 & 5.20 & -4.2 & 5.59 & 19.2 &  18\\[0.6mm]
   78   & WN7h          &  50.1 & 1.0 & 1385 & 11 & 0.47 & S     & 10.5$^{+0.2}_{-0.2}$ G  -5.83 &10.14 & -4.5 & 5.80 &  3.4 &  24 / 22\\[0.6mm]
   82   & WN7(h)        &  56.2 & 0.7 & 1100 & 20 & 1.00 & S     & 12.9$^{+0.4}_{-0.3}$ G  -4.63 & 4.24 & -4.8 & 5.26 &  4.9 &  11\\[0.6mm]
   84   & WN7 (WNE-w)   &  50.1 & 0.9 & 1100 &  0 & 1.45 & S     & 12.6$^{+0.4}_{-0.3}$ G  -4.95 & 6.30 & -4.8 & 5.36 &  3.6 &  13\\[0.6mm]
   85   & WN6h-w (WNL)  &  50.1 & 1.1 & 1400 & 40 & 0.82 & C 3.5 & 11.5$^{+0.3}_{-0.2}$ G  -4.66 & 6.46 & -5.0 & 5.38 &  3.1 &  13\\[0.6mm]

\s 87   & WN7h          &  44.7 & 1.3 & 1400 & 40 & 1.70 & S     & 12.6$^{+0.5}_{-0.4}$ g  -6.95 &20.34 & -4.5 & 6.21 &  1.3 &  44 / 59\\[0.6mm]
   89   & WN8h          &  39.8 & 1.4 & 1600 & 20 & 1.58 & S     & 12.6$^{+0.5}_{-0.4}$ G  -7.56 &30.04 & -4.4 & 6.33 &  1.5 &  53 / 87\\[0.6mm]
   91   & WN7  (WNE-s)  &  70.8 & 0.4 & 1700 &  0 & 2.12 & S     & 13.6$^{+0.7}_{-0.6}$ g  -6.11 & 6.13 & -3.9 & 5.93 & 13.5 &  29\\[0.6mm]
   94   & WN5-w         &  56.2 & 0.9 & 1300 &  - & 1.49 & C 3.4 &  9.9$^{+0.1}_{-0.1}$ G  -4.25 & 6.01 & -4.8 & 5.52 &  3.4 &  16\\[0.6mm]
   100  & WN7  (WNE-s)  &  79.4 & 0.3 & 1600 &  0 & 1.50 & S     & 13.0$^{+0.5}_{-0.4}$ G  -5.71 & 3.97 & -4.1 & 5.77 & 11.0 &  23\\[0.6mm]
   105  & WN9h          &  35.5 & 1.1 &  800 & 17 & 2.15 & S     & 11.2$^{+0.3}_{-0.3}$ g  -7.13 &23.32 & -4.4 & 5.89 &  1.9 &  27 / 25\\[0.6mm]
   107  & WN8           &  50.1 & 0.8 & 1200 &  - & 1.41 & C 3.7 & 14.6                 \l -7.22 & 16.7 & -4.0 & 6.2  &  3.9 &  44 \\[0.6mm]
   108  & WN9h          &  39.8 & 1.4 & 1170 & 27 & 1.00 & S     & 12.3$^{+0.3}_{-0.3}$ G  -6.26 &16.07 & -4.9 & 5.77 &  1.3 &  23 / 21\\[0.6mm]
   110  & WN5-s         &  70.8 & 0.5 & 2300 &  0 & 0.90 & C 3.5 & 11.0$^{+0.2}_{-0.1}$ G  -4.85 & 3.73 & -4.2 & 5.51 & 23.1 &  16\\[0.6mm]
   115  & WN6-w         &  50.1 & 0.9 & 1280 &  0 & 1.50 & S     & 11.5                 \R -5.33 & 8.89 & -4.5 & 5.65 &  4.3 &  20 \\[0.6mm]
\hline 
\end{tabular}
\newline

(to be continued)
\normalsize
\end{flushleft}
\end{table*}

\begin{table*}
\addtocounter{table}{-1}
\caption[]{(continued)}
\small
\begin{flushleft}
\begin{tabular}{ l l r r r r r l r c r l r r }
\hline\hline 
\s WR &
Spectral subtype &
$T_*$ &
$\log R_{\rm t}$ &
\multicolumn{1}{c}{$\varv_\infty$} &
$X_{\rm H}$ &
$E_{b-\varv}$ &
Law\a  &
\DM\c \hspace{5mm} $M_\varv$ &
$R_*$& log\,$\dot{M}^d$ &
log\,$L$ &
$\frac{{\dot M} \varv_{\infty}}{L/c}$ &
$M^e$ \\

&
&
[kK] &
$[R_{\odot}]$ &
\multicolumn{1}{c}{[km/s]} &
[\%] &
[mag]&
$R_V$  &
[mag]\ \ \ [mag] &
$[R_{\odot}]$ &
[$M_{\odot}$/yr] &
[$L_{\odot}$] &
&
[$M_{\odot}$] \\

   (1) & (2)           &(3)      & (4)   & \multicolumn{1}{c}{(5)}
                                             &  (6) & (7)&  (8) & (9) \hspace{0.6cm} (10)
                                                                                   & (11) & (12) & (13)& (14) &(15) \\
\hline 
\s 116  & WN8h          &  39.8 & 0.8 &  800 & 10 & 1.75 & S     & 12.0$^{+0.3}_{-0.3}$ G  -5.81 &10.97 & -4.4 & 5.44 &  5.5 &  14\\[0.6mm]
   120  & WN7 (WNE-w)   &  50.1 & 0.8 & 1225 &  0 & 1.25 & S     & 11.0$^{+0.7}_{-0.6}$ g  -3.81 & 3.78 & -4.9 & 4.92 &  8.9 &   7\\[0.6mm]
   123  & WN8 (WNE-w)   &  44.7 & 0.7 &  970 &  0 & 0.75 & C 2.8 & 13.8$^{+0.6}_{-0.5}$ G  -5.28 & 6.97 & -4.6 & 5.28 &  6.7 &  12\\[0.6mm]
   124  & WN8h          &  44.7 & 0.7 &  710 & 13 & 1.08 & C 2.9 & 14.0$^{+0.5}_{-0.4}$ G  -6.58 &11.93 & -4.3 & 5.75 &  3.2 &  22 / 20\\[0.6mm]
   127  & WN3 + O9.5V   &  \comp \\[0.6mm]
   128  & WN4(h)-w      &  70.8 & 1.1 & 2050 & 16 & 0.32 & C 3.6 & 12.3$^{+0.3}_{-0.3}$ g  -3.27 & 2.69 & -5.4 & 5.22 &  2.6 &  11\\[0.6mm]
   129  & WN4-w         &  63.1 & 0.9 & 1320 &  0 & 0.85 & S     & 13.9$^{+0.4}_{-0.4}$ G  -4.10 & 4.17 & -5.0 & 5.40 &  2.4 &  14\\[0.6mm]
   130  & WN8(h)        &  44.7 & 1.0 & 1000 & 12 & 1.46 & S     & 13.8                 \l -7.22 & 22.1 & -4.2 & 6.25 &  1.8 &  47 \\[0.6mm]
   131  & WN7h          &  44.7 & 1.3 & 1400 & 20 & 1.15 & S     & 14.5$^{+0.5}_{-0.4}$ G  -6.82 &19.12 & -4.5 & 6.14 &  1.5 &  39 / 44\\[0.6mm]
   133  & WN5 + O9I     &  \comp \\[0.6mm]

\s 134  & WN6-s         &  63.1 & 0.7 & 1700 &  0 & 0.47 & C 3.4 & 11.2$^{+0.1}_{-0.1}$ G  -5.09 & 5.25 & -4.4 & 5.61 &  8.3 &  18\\[0.6mm]
   136  & WN6(h)-s      &  70.8 & 0.5 & 1600 & 12 & 0.45 & S     & 11.4$^{+0.2}_{-0.1}$ G  -5.63 & 5.10 & -4.2 & 5.78 &  8.0 &  23 / 21\\[0.6mm]
   138  & WN5-w + B?    &  \comp \\[0.6mm]
   139  & WN5 + O6II-V  &  \comp \\[0.6mm]
   141  & WN5-w +O5V-III&  \comp \\[0.6mm]
   147  & WN8(h) + B0.5V&  39.8 & 0.9 & 1000 &  5 & 2.85 & S     & 10.4                 \l -7.22 & 29.8 & -3.8 & 6.3  &  3.6 &  51 \\[0.6mm]
   148\b& WN8h + B3IV/BH&  39.8 & 1.3 & 1000 & 15 & 0.83 & C 3.0 & 14.4                 \l -7.22 & 26.5 & -4.5 & 6.2  &  1.0 &  44 \\[0.6mm]
   149  & WN5-s         &  63.1 & 0.7 & 1300 &  0 & 1.42 & S     & 13.5$^{+0.3}_{-0.2}$ g  -4.65 & 4.27 & -4.6 & 5.43 &  5.5 &  14\\[0.6mm]
   151  & WN4 + O5V     &  \comp \\[0.6mm]
   152  & WN3(h)-w      &  79.4 & 1.1 & 2000 & 13 & 0.50 & C 3.2 & 13.3$^{+0.5}_{-0.4}$ G  -3.74 & 3.63 & -5.2 & 5.68 &  1.4 &  20 / 18\\[0.6mm]

\s 155  & WN6 + O9II-Ib &  \comp \\[0.6mm]
   156  & WN8h          &  39.8 & 1.1 &  660 & 27 & 1.22 & S     & 13.1$^{+0.2}_{-0.2}$ G  -7.00 &20.81 & -4.6 & 6.01 &  0.9 &  32 / 32\\[0.6mm]
   157  & WN5-w (+B1II) &  \comp \\[0.6mm]
   158  & WN7h + Be?    &  44.7 & 1.2 &  900 & 30 & 1.08 & S     & 13.6$^{+0.3}_{-0.3}$ G  -6.49 &17.85 & -4.7 & 6.06 &  0.7 &  35 / 35\\[0.6mm]
\hline 
\end{tabular}
\newline

\begin{itemize}
\item[\a] Applied reddening law: S = Seaton \citep{Seaton1979},
C = Cardelli et al. \citep{Cardelli+al1989}, F = Fitzpatrick \citep{Fitzpatrick1999};
for the last two, the given number is the adopted $R_V$

\item[\b]  Binary system in which the non-WR component contributes
more than 15\% of the flux in the visual

\item[\c]  A letter ``G'' following this column indicates that the
distance is based on a Gaia parallax; a lower case ``g'' indicates that
this measurement is flagged because of significant {\em astrometric
excess noise}

\item[$^d$]  Mass-loss rates are for an adopted clumping factor of $D$ = 4

\item[$^e$]  Current stellar mass from an $M$-$L$ relation for homogeneous
helium stars; if a second value is given, the latter is derived from an
$M$-$L$ relation for WNL stars based on evolutionary tracks (see text)

\end{itemize}

\normalsize
\end{flushleft}
\end{table*}


Comparing the Gaia distances for our remaining sample (55 stars) with those
adopted in Paper\,I reveals a mild tendency to smaller values. The
arithmetic mean over all distance moduli became lower by 0.2\,mag, which
corresponds to 10\% smaller distances on average. However, for certain
WN stars the \DM\ values differ significantly. Revisions of the distance 
modulus by more than 1.5\,mag are encountered for WR\,16
($\Delta\DM$=1.5\,mag), WR\,82 (-2.6\,mag), WR\,83 (-2.6\,mag), WR\,120
(-1.7\,mag), and WR\,123 (-1.9\,mag). 

\begin{figure}[!tb]
\centering
\epsfxsize=\columnwidth
\mbox{\epsffile{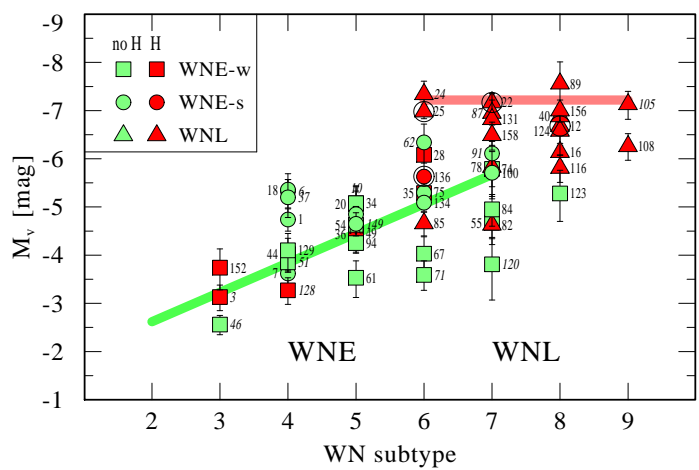}}
\caption{Absolute narrow-band $\mathrm{\varv}$-magnitudes of Galactic 
WN stars based  on Gaia DR2 distances. The red symbols refer to stars
with detectable hydrogen, while the green filled symbols stand for
hydrogen-free stars. The symbol shapes refer to the spectral subtype:
``early'' and ``late'' subtypes (WNE and WNL, respectively) and
distinguish among the WNE stars between weak (-w) and strong (-s)
lines (see Paper\,I for details). The binaries and runaways are encircled (cf.\
Sect.\,\ref{sect:binarity}). The labels refer to the WR catalog numbers;
if printed with slanted font, their Gaia measurement is
flagged with a significant astrometric excess noise (see
Sect.\,\ref{sect:distances}).  The thick lines indicate the subtype
calibration adopted in Paper\,I. 
}
\label{fig:mvcalib}
\end{figure}

\subsection{$M_\mathrm{\varv}$-WN subtype calibration}

Photometry for WR stars is preferably considered in terms of the
narrowband magnitudes defined by \cite{Smith1968}, using lower case
letter subscripts (e.g., $M_\mathrm{\varv}$, $M_\mathrm{b}$) to
distinguish them from Johnson broad-band colors. Adopting the same
apparent magnitudes as in Paper\,I, and also keeping for each star 
the same reddening law and the same value for the color excess as in Paper\,I, 
the absolute visual magnitude $M_\mathrm{\varv}$ changes just by
$\Delta\DM$ due to the Gaia revision of the distance modulus. 
  
The resulting absolute narrowband $\mathrm{\varv}$ magnitudes of  our
sample stars  are plotted in Fig.\,\ref{fig:mvcalib} versus their WN
subtype.  In Paper\,I  we had to rely on the assumption that there are
strict correlations (the red and green thick lines in 
Fig.\,\ref{fig:mvcalib} for the WNL and WNE stars, respectively), and
used these relations to predict the absolute magnitudes for the
majority of the sample stars for which no other distance estimate
existed.  

Based on the  Gaia measurements,  Fig.\,\ref{fig:mvcalib} reveals
that the correlations between subtype and absolute visual magnitude are by no
means strict. If, for instance, all stars of subtype WN4 (without
hydrogen)  were to have the same absolute magnitude, two-thirds of the
green symbols in the WN4  Column (i.e., five out of eight) would overlap with
this true value within their $1\sigma$ error bars. This is obviously
not the case. For the WN stars with hydrogen (red symbols in
Fig.\,\ref{fig:mvcalib}) the Gaia distances confirm their generally
high brightness, but also here the scatter (e.g., for the WN8 subtype)
is larger than expected from the statistical error of the distance. 
Thus we must conclude that a specific WN subtype can be reached by
stars with different mass, luminosity, and history.

\section{Rescaling of the stellar parameters}
\label{sect:parameters}

The spectroscopic parameters, which are not affected by the adopted
distance, are repeated from Paper\,I in Columns 3--8 of 
Table\,\ref{table:parameters}. One of these spectroscopic parameters is 
the {\em transformed radius}, i.e.,
  \begin{equation}
  \label{eq:rt}
 R_{\mathrm{t}} = R_{\ast} \left[ \frac{\varv_{\infty}}{2500\,\mathrm{km/s}} 
 \left/ \frac{\dot{M} \sqrt{D} }{10^{-4} M_{\odot}/\mathrm{yr}} \right. 
 \right]^{\frac{2}{3}} \quad .
  \end{equation} 

In this   equation, $R_\ast$ denotes the stellar radius that  corresponds, by our
definition, to a Rosseland continuum optical depth of $\tau = 20$;
$\dot{M}$ denotes the mass-loss rate; $\varv_\infty$ is the terminal wind
velocity; and $D$ is the clumping factor as introduced in
\citet{HK1998}.         

The (misleadingly termed) ``transformed radius'' has been defined  
\citep{SHW1989,HK1998} when realizing that normalized line spectra 
for WR stars of same $T_\ast$ depend only on $R_{\mathrm{t}}$, but are
nearly independent from the individual combination of the other
parameters (especially, $R_\ast$ and $\dot{M}$) that enter
Eq.\,(\ref{eq:rt}). Thus, the use of $R_{\mathrm{t}}$ reduces the
dimension of the parameter space for which  models must be provided;
one can fit the normalized line spectrum with models for a ``wrong''
luminosity, and afterwards re-scale the spectral energy distribution
according the observed flux (see Paper\,I for more explanations).   

For a spherically extended object, the definition of an effective 
temperature depends on the reference radius. The {\em stellar
temperature} $T_\ast$ (Column\,3) refers to the stellar radius $R_\ast$ 
defined above. Thus, the Stefan-Boltzmann equation 
\begin{equation}
\label{eq:stefan-boltzmann}
L = 4 \pi R_\ast^2 ~\sigma_\text{SB} T_\ast^4
\end{equation}
relates $R_\ast$ and the corresponding (effective) stellar temperature 
$T_\ast$ to the stellar luminosity $L$. 

By combining Eqs.\,(\ref{eq:rt}-\ref{eq:stefan-boltzmann}) one 
can immediately obtain the corrections to the stellar luminosity, 
radius, and mass-loss rate that follow from a revision of the distance
modulus by $\Delta\DM$:
\begin{align}
\label{eq:scalings}
\Delta\log L      &= 0.4 \cdot \Delta\DM \quad ,\\
\Delta\log R_\ast &= 0.2 \cdot \Delta\DM \quad ,\\
\Delta\log\dot{M} &= 0.3 \cdot \Delta\DM \quad .
\end{align}

After being rescaled to the Gaia distances,  the stellar parameters
($R_\ast$, $\dot{M}$, $L$) are compiled in
Table\,\ref{table:parameters}, Columns (11), (12), and (13),
respectively. In Column\,(14) we recalculate the ``wind efficiency''
$\eta = \dot{M} \varv_\infty/(L/c)$,  i.e., the ratio of the
mechanical momentum of the wind to  the radial momentum in the
radiation field (per unit of time). 

The last Column\,(15) gives current stellar masses. These values are
calculated from the mass-luminosity relation for helium-burning stars on
the helium zero-age main sequence taken from \cite{Graefener+2011}.
This approximation  might actually not be adequate for all
stars in our sample, especially not if hydrogen is still detected in
their atmosphere. For all stars with $X_{\rm H} > 0$ and $\log
L/L_\odot > 5.5$ we construct a special $M-L$ relation for the WNL
stage based on the evolutionary tracks for rotating single stars from
\cite{ekstroem+2012} -- see Fig.\,\ref{fig:hrd+tracks}. The stellar
mass as derived from that relation is given as second value in
Column\,(15). We note that the stellar masses are significantly larger
only for very high luminosities ($\log L/L_\odot \gtrsim 6.2$) for
which the tracks   predict ongoing hydrogen burning in the core. 

\section{Results and discussion}

\subsection{Mass-loss rate versus luminosity}

\begin{figure}[!tb]
\centering
\epsfxsize=\columnwidth
\mbox{\epsffile{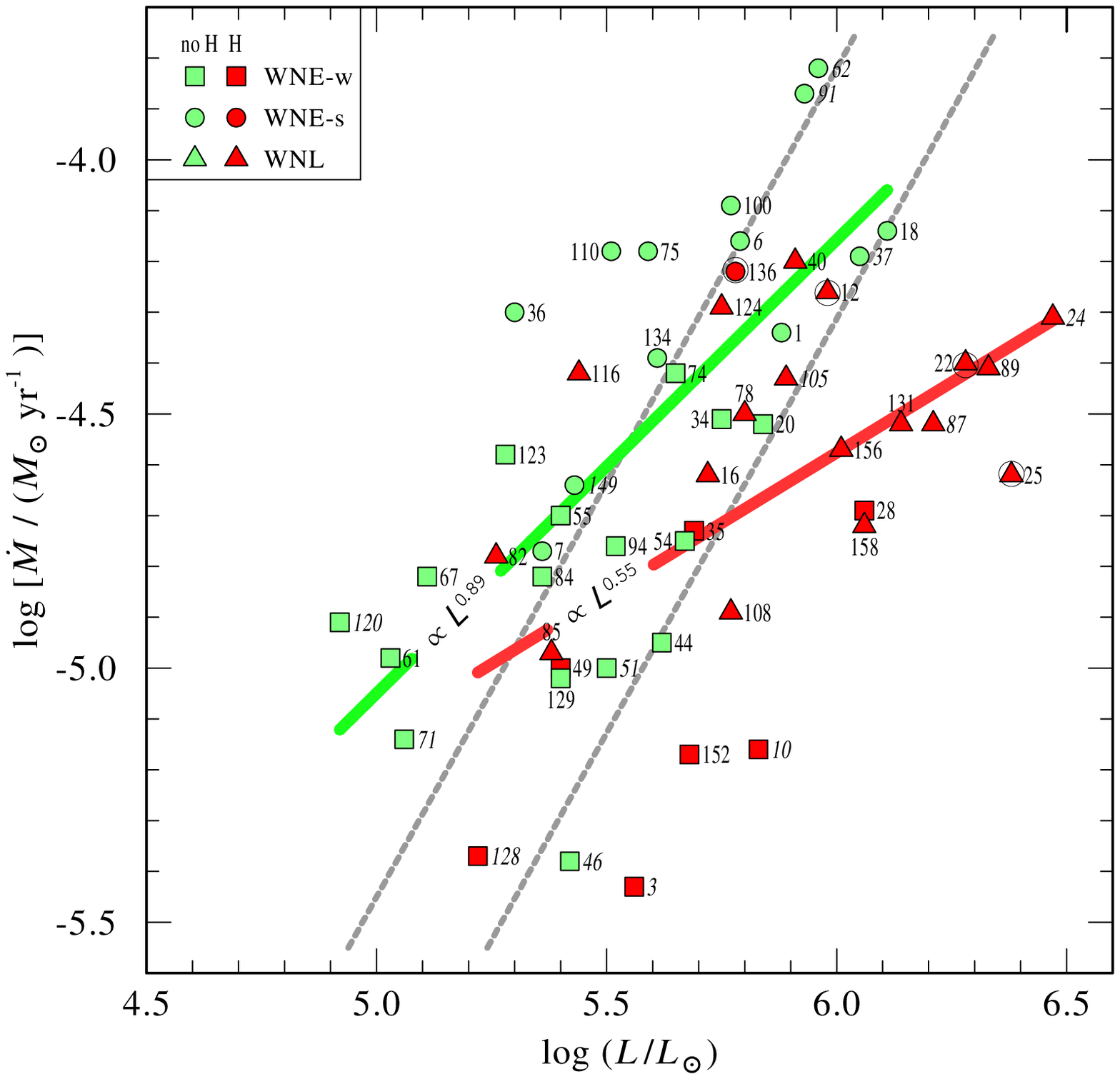}}
\caption{Empirical mass-loss rate versus luminosity for the Galactic WN
stars. Labels refer to the WR catalog numbers;
if printed with slanted font, their Gaia measurement is
flagged with a significant astrometric excess noise (see
Sect.\,\ref{sect:distances}). The red symbols refer to stars with detectable
hydrogen, while green symbols denote hydrogen-free
stars (see inlet and caption of Fig.\,\ref{fig:mvcalib} for more details). Neither group
follows a tight $\dot{M}-L-$relation. The thick green and red lines are linear
regressions to the stars without and with hydrogen, respectively. 
The gray dotted lines denote the corresponding empirical relations 
from \cite{NugisLamers2000} for hydrogen-free WN stars (upper line) and
for a hydrogen mass fraction of 40\% (lower line).
}
\label{fig:mdot}
\end{figure}

\begin{figure*}[!t]
\centering
\epsfxsize=0.60\textwidth
\mbox{\epsffile{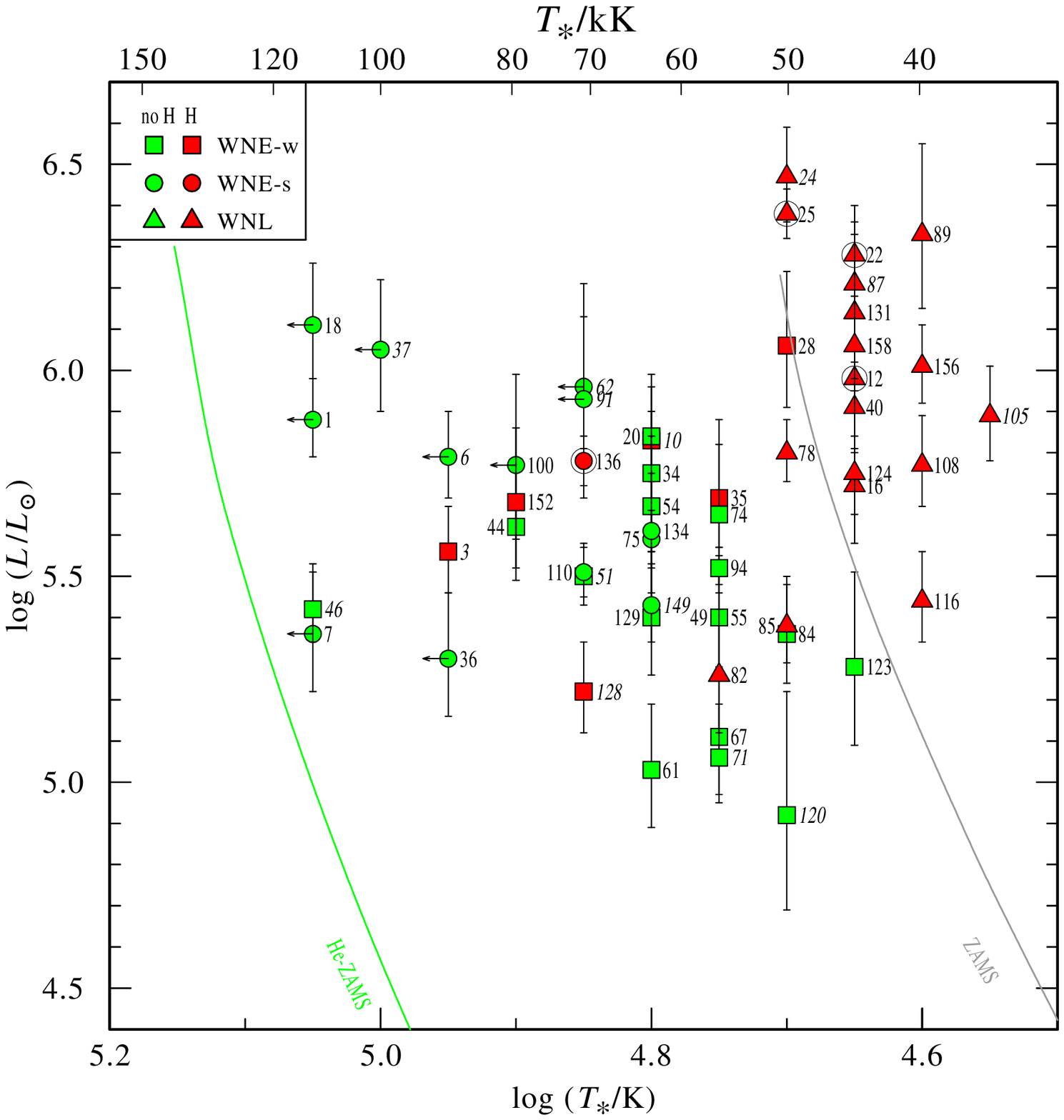}}
\caption{Hertzsprung-Russell diagram of Galactic WN stars with Gaia
parallax. The filling color reflects the surface composition (red:
with hydrogen; green: hydrogen-free), while the symbol shape refers to
the spectral subtype (see caption of Fig.\,\ref{fig:mvcalib}). The binaries
and runaways are encircled.  The labels refer to the WR catalog numbers;
if printed with slanted font, their Gaia measurement is
flagged with a significant astrometric excess noise (see
Sect.\,\ref{sect:distances}).  A little arrow indicates that for this
particular star the indicated temperature is only a lower limit,
because of the parameter degeneracy discussed in the text
(Sect.\,\ref{sect:hrd}). The vertical error bars reflect only the
$1\sigma$ margin of the Gaia distances. Just for orientation, the Zero
Age Main Sequences are indicated for solar composition and for pure
helium stars, respectively.  }
\label{fig:hrd-wn}
\end{figure*}

Radiation-driven mass loss is expected to depend, in first place, on
the stellar luminosity. With the use of Gaia parallaxes, the stellar 
luminosities became more reliable. Therefore, we expected that the 
$\dot{M}-L-$plot (Fig.\,\ref{fig:mdot}) would show a better defined
relation than obtained previously (cf.\ Paper\,I). However, the contrary is the
case. The hydrogen-free stars (green symbols) show at least some 
correlation; the linear regression yields 
\begin{equation}
\log \dot{M} = 
0.89 \times \left( \log(L/L_\odot) - 4.92 \right) - 5.12
~[M_\odot/\mathrm{yr}]
,\end{equation}
where the formal error of the slope is $\pm 0.18$. 
The distribution of the red symbols (stars with hydrogen) looks even
more messy. Obviously, this group of stars is by no means uniform; it
contains stars in very different evolutionary stages and with
quantitatively different hydrogen mass fraction in their winds.  
Formally, the linear regression to the red symbols yields
\begin{equation}
\log \dot{M} = 
0.55 \times \left( \log(L/L_\odot) - 5.22 \right) - 5.01 
~[M_\odot/\mathrm{yr}]
,\end{equation}
where the formal error of the slope is $\pm 0.17$. 

The slopes of these regression lines are shallower than previously
found. Figure\,\ref{fig:mdot} also shows the older empirical relations
claimed by \cite{NugisLamers2000} that have a slope of 1.63 (gray dotted
lines). 
However the shallow slopes found here resemble the value   of 
$0.68\pm 0.05$ obtained recently for the Galactic WC stars by 
\cite{Sander+2019} using the new Gaia distances. 
Shallow slopes are in
line with theoretical expectations from hydrodynamical models
especially for the hydrogen-free stars  because of their close
proximity to the Eddington limit and the physics of wind-driving
\citep{Graefener+Hamann2008, Graefener+2011}. For optically thick winds,
\cite{Graefener+2017} predicted a slope of 1.3. 

However, the empirical correlations shown in Fig.\,\ref{fig:mdot}
are by no means tight, but have a large scatter. One might speculate
that further parameters play a role, for example, different iron abundance or
variations of the clumping properties. Since such parameters are not yet
at hand, we refrain  from a further discussion of the $\dot{M}$
dependencies. Closer studies of these questions would be interesting.  

\subsection{Hertzsprung-Russell diagram}
\label{sect:hrd}

The empirical HRD of our Galactic WN
sample is shown in Fig.\,\ref{fig:hrd-wn}. As in the corresponding HRD
in Paper\,I, the locations of the hydrogen-containing WNL stars and of 
the hydrogen-free WNE stars are found to be divided by the hydrogen
Zero Age Main Sequence (ZAMS). Only a few WNL and WNE stars with
hydrogen (red symbols in Fig.\,\ref{fig:hrd-wn}) violate this general
rule.  However, now using the Gaia distances, the WNL and WNE
stars are not separated by a gap in luminosities anymore. Instead, in
the range between $\log L/L_\odot$ = 5.7 and 6.1 both  subclasses can
be found.  This is in line with the population synthesis presented in
Paper\,I. 

We must remark on those symbols in the HRD that bear a
little arrow to the left.   As discussed in Paper\,I, there are a few
stars in the sample that have such a thick wind that the entire
spectrum is formed far out  in  the wind. In other words, the
pseudo-photosphere expands with a significant fraction of the terminal
velocity, while any quasi-static layers are deeply embedded and cannot
be observed. Having adopted a $\beta$-law for the velocity field (cf.\
Paper\,I) throughout the supersonic part of the wind, we implicitly
assume a steep velocity gradient for the inner wind region, and thus a
stellar radius $R_\ast$ that is only slightly smaller than the radius
from where the radiation emerges. However, if the velocity law in the
lower, unobservable part of the wind would be in fact shallow, or if
there was any other form of an ``inflated envelope'', the quasi-static
stellar ``core'' might have a significantly smaller radius than the
pseudo-photosphere, thus potentially implying a much higher ``stellar
temperature'' than the $T_\ast$ obtained by our analysis (cf.\
Eq.\,\ref{eq:stefan-boltzmann}).  For such a thick atmosphere, the
normalized line spectrum depends to first order only on the 
ratio $L/\dot{M}^{4/3}$, but not on $T_\ast$ and $R_\ast$. 

\begin{figure}[!t]
\centering
\epsfxsize=1.0\columnwidth
\mbox{\epsffile{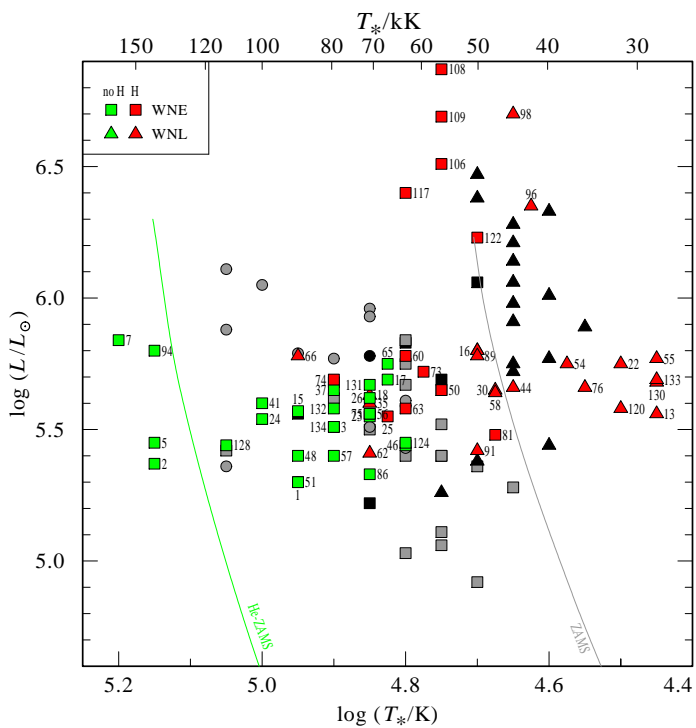}}
\caption{Hertzsprung-Russell diagram of the WN stars in the LMC
\citep[adapted from][]{Hainich+2014}. The filling color 
reflects the surface composition (red:
with hydrogen; green: hydrogen-free), while the symbol shapes refer to
the spectral subtype (see inlet). The labels refer to the
BAT\,99 catalog numbers \citep{BAT99}. 
For comparison, the Galactic WN stars studied in the present paper are 
represented by the grey (hydrogen-free) and black symbols in the
background. For orientation, the Zero
Age Main Sequences are indicated for solar composition and for pure
helium stars, respectively.  }
\label{fig:lmc-hrd}
\end{figure}

It is interesting to compare our empirical HRD for the Galactic WN stars
with the corresponding HRD for single WN stars in the Large Magellanic 
Cloud (LMC), which had been analyzed in a similar way by \citet{Hainich+2014} 
while distance uncertainties are not an issue for LMC members. 
Differences between the Galactic and the LMC sample can, in
principle, have two reasons: the different metalicities, and a
different star formation history.

Fig.\,\ref{fig:lmc-hrd} reveals that the distributions of both samples
are generally similar, but differ in detail. The LMC contains
a few very luminous stars 
\citep[BAT99\,106, 108 and 109,][]{Crowther+2010}
These stars all reside in
the very active 30\,Dor Starburst complex. Similarly, the most luminous
members of our Galactic sample are preferably found in the Carina
nebula, and thus also in a massive star-forming region. Obviously, such
an environment is favorable for finding very massive stars. 

The luminosity distribution of the LMC sample shows a pronounced gap
between the seven most luminous stars and the numerous rest. A similar
luminosity gap between the WNL and the WNE stars was originally visible
in Paper\,I for the Galaxy as well,  but is now filled with WNE and WNL
stars as the result of the Gaia distances. Thus the bimodal
distribution of WN luminosities found in Paper\,I was an artifact
introduced by the subtype calibration of absolute magnitudes, but in the
LMC it is real and probably reflects a particular age distribution.  

At the lower end of the luminosity distribution, we find a couple of
Galactic WNE stars, but no LMC counterparts. This can be explained in
terms of single-star evolution by the metallicity dependence of
stellar wind mass loss. Because of the lower metallicity in the LMC, the
minimum mass required to bring an evolutionary track back to the hot
side of the HRD is somewhat higher in the LMC than in the Galaxy
\citep[e.g.,][]{Meynet+2005}. If the WNE stars have lost their
hydrogen envelope by Roche-lobe overflow (RLOF) in close-binary evolution, a
metallicity dependence is not expected, and the observed difference
between those two samples would have no obvious explanation. 

Regarding the temperature distribution, the WN stars in the LMC appear
to be slightly hotter than their Galactic counterparts; there are a
few WNE stars residing even to the left of the helium ZAMS, and many WN
stars with hydrogen are hotter than the hydrogen ZAMS. This could again
indicate a metallicity effect. Since the LMC stars have slightly more
compact cores and less thick winds, their spectra show higher
stellar temperatures.

\begin{figure}[!t]
\centering
\epsfxsize=\columnwidth
\mbox{\epsffile{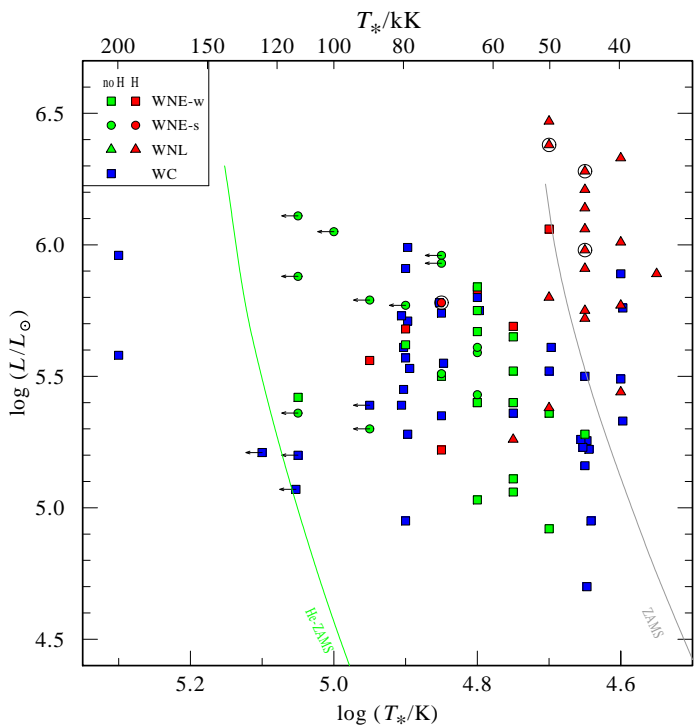}}
\caption{Empirical HRD of the Galactic WN sample
(green and red symbols, cf.\ Fig.\,\ref{fig:hrd-wn}), now additionally 
including Galactic WC and WO-type 
stars (blue symbols) from \cite{Sander+2019}.} 
\label{fig:hrd-wn+wc}
\end{figure}

In Fig.\,\ref{fig:hrd-wn+wc} we compare the HRD positions of the
Galactic WN sample studied in this paper with those of the Galactic WC
and WO stars \citep{Sander+2019}, which  have also been updated
according to the Gaia DR2 distances. 
The WC and WO stars must have evolved from hydrogen-free WN stars, 
when the latter have
lost their helium layers and display the products of helium burning in
their atmosphere. As Fig.\,\ref{fig:hrd-wn+wc}  shows, the WC stars
populate a similar range in luminosities as the WN stars with a slight
tendency toward lower values, which is expected due to the progressing loss
of mass. However, many WC stars have lower stellar
temperatures than the WNE stars. Obviously, the evolution WNE
$\rightarrow$ WC does not proceed toward higher $T_\ast$.  
We conclude that the outer layers of a star become more
``inflated'' \citep{Graefener+2012} at the transition from the WNE to 
the WC stage.

\subsection{Stellar evolution}

\begin{figure}[!t]
\centering
\epsfxsize=\columnwidth
\mbox{\epsffile{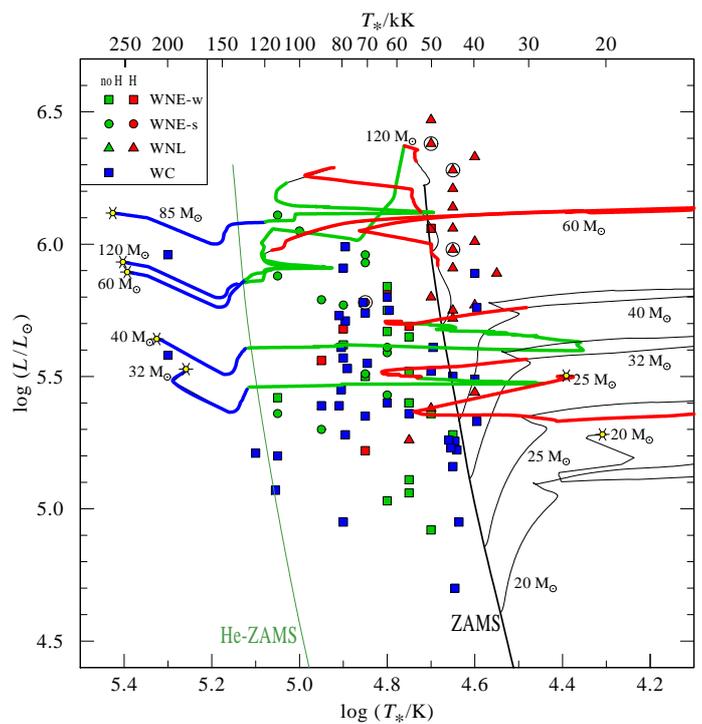}}
\caption{Hertzsprung-Russell diagram of the Galactic Wolf-Rayet stars
with Gaia distances (cf.\ Fig.\,\ref{fig:hrd-wn+wc})
compared to evolutionary tracks for single stars from
\cite{ekstroem+2012} which account for rotation. The tracks are colored
analogously in the different WR stages, according to their predicted
surface composition. The labels refer to the initial mass. } 
\label{fig:hrd+tracks}
\end{figure}

Various groups have published evolutionary tracks for massive stars with
Galactic metallicity. A thorough comparison of our results with those
various model predictions is beyond the scope of the present paper. In
Fig.\,\ref{fig:hrd+tracks} we employ the Geneva tracks for Galactic 
metallicity ($Z=0.14$) from \cite{ekstroem+2012}, which account for 
rotationally induced mixing under the assumption that all stars rotate 
initially with 40\% of their break-up rate. 

At highest luminosities, the tracks (for initially 85\,$M_\odot$ and
120\,$M_\odot$)  evolve from the hydrogen ZAMS immediately to the left,
as expected for almost homogeneous stars, while the observed  stars are
in fact located slightly above or to the right of the ZAMS. It is not clear if
this discrepancy could be attributed to envelope inflation. Such
evolutionary models show core hydrogen burning and can in principle
account for the more luminous members of the WNL stars observed. 

The track for initially  60\,$M_\odot$ shows an unrealistic excursion
to the red side, 
violating the Humphreys-Davidson limit.
Apart from this, the lower half of the luminosity
distribution of the hydrogen-displaying WNL stars can in principle be
explained by tracks that return from the cool side, i.e., after having
ignited core helium burning. Such tracks could then continue to the
hotter, hydrogen-free WNE stars which now, with the revised
distances, are found at similar luminosities. However, it seems to be a
strange coincidence (and is not predicted in detail by the tracks) why
the transition to hydrogen-free atmospheres should happen just when
the track crosses a division line that apparently coincides with the
hydrogen ZAMS.    
The location of the final WC/WO stage is in agreement with the
observed locations of the Galactic WO stars WR\,102 and WR\,142 at
200\,kK. 

In the intermediate luminosity range, WN stars with and without hydrogen
as well as WC stars are predicted by the tracks. The stellar
temperatures are discrepant, which must be attributed to envelope
inflation that is not reproduced by the evolutionary  models. As
already discussed above, especially the relatively cool temperatures of
many WC stars imply that the evolution from the WNE to the WC stage is
accompanied by a redward loop of the evolutionary track. 

A significant number of WN stars in our sample and among the
Galactic WC stars are found to have low luminosities that are not
reproduced by the evolutionary tracks.  The lowest mass for which the
track reaches the WR stage is for  initially $25\,M_\odot$ (see
Fig.\,\ref{fig:hrd+tracks}), and even this star is predicted to
explode  before all hydrogen has been removed, i.e., it does not reach
the hydrogen-free WNE and WC stages. The track for initially
$20\,M_\odot$ stays hydrogen-rich till the end.  This could indicate
that single-star evolution models still suffer from some deficiencies,
possibly regarding the pre-WR mass-loss rates, rotational and other
mixing processes, or core overshooting. 

An alternative channel to produce WR stars invokes mass transfer in
binary systems \citep{Paczynski1967}. It has long been known that the
majority of massive stars have been born in close binary systems
\citep[e.g.,][]{Vanbeveren+1998}. 

If the primary star of a close binary falls in the suitable mass range
(i.e.,\ below $60\,M_\odot$ according to the tracks in 
Fig.\,\ref{fig:hrd+tracks}), it will expand when evolving  until
RLOF occurs. After its hydrogen envelope is
removed (and partially accreted on the companion),  the remaining star
is expected to be hot and hydrogen depleted, thus retraining the
spectral appearance of a WR star. Since this channel does not depend
on wind mass loss, it can work at lower luminosities than the single-star
evolution. 

However, the secondary in this system would gain a lot of mass from
RLOF, and thus would become most likely a bright and detectable OB
star.  For our sample stars, a currently present OB-type companion is
observationally excluded (except for the few established binaries, see
Sect.\,\ref{sect:binarity}).  Thus, such history cannot explain the
apparently single WN stars of our sample.

As a theoretical possibility to avoid a bright companion, common
envelope evolution has been proposed. If the secondary was originally a
low-mass star (e.g., with 1\,$M_\odot$), it might have helped to eject the
hydrogen envelope without accreting much mass itself
\citep[e.g.,][]{Kruckow+2016}, albeit there are energetic constraints to
the possible parameters of such a scenario. Finally, the components might 
have merged. If it survived, such faint, low-mass companion might
be very difficult to detect. 

Alternatively, we may consider the possibility that the current WR
star was originally the secondary of a binary system and served as
the mass gainer in a first RLOF. Then the primary explodes or collapses to a
compact object. If the system stayed bound after this event, a
second RLOF phase could occur, this time from the original secondary
to the compact object. This could help to strip off the hydrogen
envelope from the secondary and turn it into a WR star. This channel
would lead to a WR + compact companion system, which would be easily
detected as a High Mass X-ray Binary, and therefore must also be ruled
out for our sample -- unless the compact companion is so deeply
embedded in the WR wind that X-rays cannot emerge, as has been
speculated recently for WR\,124 \citep{Toala+2018}.  Even more exotic,
the compact companion might have been engulfed in a merger process
forming a kind of Thorne-\.Zytkow object, as has been occasionally 
speculated to be the nature of WN8 stars
\citep[e.g.,][]{Foellmi+Moffat2002}. 

Summarizing the discussion of the HRD, it seems that neither the
single-star evolutionary models considered in this work, nor binary scenarios can
provide a fully satisfactory explanation for the parameter distribution
of the apparently single Galactic WN stars.  Most likely,  the
single-star evolutionary calculations still suffer from incomplete
physics, such as mass-loss recipes, internal mixing, and envelope
inflation.

\section{Summary}
\label{sect:summary}

Spectral analyses of a comprehensive set of Galactic WN stars, mostly
putatively single, have been presented by \cite{Hamann+2006} (Paper\,I).
These analyses were based on the comparison with synthetic spectra 
calculated with the Potsdam Wolf-Rayet (PoWR) non-LTE stellar atmosphere 
code. 

\begin{enumerate}

\item
At the time of Paper\,I, the distances to the individual objects in this
Galactic sample were poorly known. The distance uncertainty affects the
``absolute'' quantities derived from the spectral analysis, such as
luminosity and mass-loss rate. 

\item
Only recently, trigonometric parallaxes became
available for the first time from the Gaia satellite (DR\,2) 
for nearly all objects in this sample (now 55 objects). On 
average, the new distances are smaller by only 10\% compared to the
values adopted in Paper\,I. However, for some of the objects the revisions
are substantial (-2.6\,mag in distance modulus in the two most extreme
cases).  

\item 
In this work, we keep the spectroscopic parameters from the analyses in
Paper\,I, but rescale the results according to the new distances based
on the Gaia parallaxes. 

\item
The correlations between mass-loss rate and luminosity show a large
scatter, for the hydrogen-free WN stars as well as for those with
detectable hydrogen. The slopes of the $\log L - \log \dot{M}$
correlations are shallower than found previously. 

\item
The empirical HRD still shows the previously
established dichotomy between the hydrogen-free early WN subtypes, which
are located on the hot side of the ZAMS, and the 
late WN subtypes, which show hydrogen and reside mostly at cooler
temperatures than the ZAMS (with few exceptions). 

\item
With the new distances, the distribution of stellar luminosities became 
more continuous than obtained previously. The hydrogen-showing WNL stars
are still found to be typically more luminous than the hydrogen-free
WNEs, but there is a range of luminosities ($\log L/L_\odot \approx$
5.5 ... 6.1) where both subclasses overlap. 

\item
The empirical HRD of the Galactic single WN stars is compared with
recent evolutionary tracks from \cite{ekstroem+2012}. 
Neither these single-star evolutionary models
nor binary scenarios can provide a fully satisfactory explanation for
the parameters of these objects and their location in the HRD. 

\end{enumerate}

\begin{acknowledgements}

We thank the referee, P.\,Crowther, for his useful comments. 
This work has made use of data from the European Space Agency (ESA)
mission {\it Gaia} (\url{http://www.cosmos.esa.int/gaia}), processed by
the {\it Gaia} Data Processing and Analysis Consortium (DPAC;
\url{http://www.cosmos.esa.int/web/gaia/dpac/consortium}). Funding for
the DPAC has been provided by national institutions, in particular the
institutions participating in the {\it Gaia} Multilateral Agreement.
R.H.\  was supported by the German \emph{Deut\-sche
For\-schungs\-ge\-mein\-schaft} (DFG) project HA\,1455/28-1.
A.A.C.S.\ was supported by the DFG project 
HA\,1455-26, and also thanks the STFC for funding
under grant ST/R000565/1.
T.S.\ acknowledges funding from the German ``Verbundforschung'' (DLR)
grant 50\,OR\,1612 and from the European  Research Council (ERC) under the
European Union's {\sc dlv\_772225\_multiples} Horizon 2020 research and
innovation programme. L.M.O acknowledges support from the DLR under
grant 50\,OR\,1809 and partial support from the Russian Government
Program of Competitive Growth of the Kazan Federal University.

\end{acknowledgements}

\bibliographystyle{aa} 
\bibliography{galwngaia}

\end{document}